\documentclass[%
 prx,
superscriptaddress,
twocolumn,
 amsmath,amssymb,
 aps,
]{revtex4-2}
\usepackage{graphicx}% Include figure files
\usepackage{dcolumn}% Align table columns on decimal point
\usepackage{bm}% bold math
\usepackage[colorlinks,linkcolor=blue,urlcolor=blue,citecolor=blue]{hyperref}
\usepackage{comment}
\usepackage{color}
\newcommand{\bs}[1]{{\boldsymbol{#1}}}

\begin{document}

\preprint{APS/123-QED}

\title{Coarsening dynamics of Ising-nematic order in a frustrated Heisenberg antiferromagnet}

\author{Yang Yang}
\affiliation{Department of Physics, University of Virginia, Charlottesville, VA 22904, USA}

\author{Yi-Hsuan Liu}
\affiliation{Department of Physics, University of Virginia, Charlottesville, VA 22904, USA}
\affiliation{Department of Physics, National Tsing Hua University, Hsinchu 30013, Taiwan}

\author{Rafael M. Fernandes}
\affiliation{Department of Physics, University of Illinois Urbana-Champaign, Urbana, IL 61801,
USA}
\affiliation{Anthony J. Leggett Institute for Condensed Matter Theory, University of Illinois Urbana-Champaign, Urbana, IL 61801,
USA}

\author{Gia-Wei Chern}
\affiliation{Department of Physics, University of Virginia, Charlottesville, VA 22904, USA}

\date{\today}

\begin{abstract}
    We study the phase ordering dynamics of the classical antiferromagnetic $J_1$--$J_2$ (nearest-neighbor and next-nearest-neighbor couplings) Heisenberg model on the square lattice in the strong frustration regime ($J_2/J_1 > 1/2$). While thermal fluctuations preclude any long-range magnetic order at finite temperatures, the system exhibits a long-range spin-driven nematic phase at low temperatures. The transition into the nematic phase is further shown to belong to the two-dimensional Ising universality class based on the critical exponents near the phase transition. Our large-scale stochastic Landau-Lifshitz-Gilbert simulations find a two-stage phase ordering when the system is quenched from a high-temperature paramagnetic state into the nematic phase. In the early stage, collinear alignments of spins lead to a locally saturated Ising-nematic order. Once domains of well-defined Ising order are developed, the late-stage relaxation is dominated by curvature-driven domain coarsening, as described by the Allen-Cahn equation. The characteristic size of Ising-nematic domains scales as the square root of time, similar to the kinetic Ising model described by the time-dependent Ginzburg-Landau theory. Our results confirm that the late-stage ordering kinetics of the spin-driven nematic, which is a vestigial order of the frustrated Heisenberg model, belongs to the dynamical universality class of a non-conserved Ising order. Interestingly, the system shows no violation of the superuniversality hypothesis under weak bond disorder. The dynamic scaling invariance is preserved in the presence of weak bond disorder. We also discuss possible applications of our results to materials for which vestigial Ising-nematic order is realized.
\end{abstract}

\maketitle

\section{Introduction}
    The antiferromagnetic $J_1$--$J_2$ (nearest and next-nearest neighbor interactions) Heisenberg model on a square lattice has been extensively studied over the last few decades due to its simplicity in capturing the essential physics of frustrated magnetism \cite{Misguich2005,Chandra1988,Dagotto1989,Gelfand1989,Chandra1990,Bishop1998,Kotov1999,Weber2003,Mambrini2006,Darradi2008,Jiang2009,Richter2010,Gauthe2022,Jiang2023} and its relevance to high-temperature iron-based superconducting materials \cite{Si2008,Seo2008,Xu2008,Fang2008,Fernandes2010,daConceicao2011,Fernandes2012,Liang2013}. The classical phase diagram at zero temperature gives two types of magnetically ordered ground states with respect to the ratio of $J_2/J_1$. When $J_2/J_1<0.5$, we have the well-known N\'eel order characterized by the wave vector $\mathbf{Q}=(\pi,\pi)$. When $J_2/J_1>0.5$, the ground state consists of two independent copies of the N\'eel orders on two sublattices partitioned from the square lattice, forming a stripe magnetic configuration \cite{Chandra1988}. While thermal fluctuations prevent spins from breaking the continuous rotational symmetry to form long-range magnetic order at any finite temperature due to Mermin-Wagner theorem, an order-by-disorder mechanism \cite{Villain1980,Henley1989} selects the ground state in which the two Néel orders on the sublattices are collinearly aligned when $J_2/J_1>0.5$. This alignment gives rise to a discrete Ising degree of freedom corresponding to a $\mathbb{Z}_2$ symmetry that can be broken at finite temperatures, leading to an Ising transition \cite{Chandra1990}.

    This Ising symmetry is related to the lowering of the fourfold rotational symmetry of the square lattice to twofold, as the collinear alignment between the two N\'eel vectors implies that the spins will be locked parallel to each other along one bond direction but antiparallel to each other along the orthogonal bond direction. Thus, the Ising transition is dubbed an Ising-nematic transition \cite{Fang2008,Xu2008,Fernandes2012}, since rotational symmetry is broken whereas translation symmetry is preserved \cite{Kivelson98}. The corresponding Ising-ordered state is thus denoted a vestigial phase of the magnetic stripe state \cite{Fernandes2019}. This framework has been successfully invoked to explain the nematic phase observed in the iron-pnictide superconductors \cite{Chu2010,Chuang2010,Yi2011,Chu2012,Bohmer2014,Kuo16,Fernandes2022}, whose phase diagrams often show a tetragonal-to-orthorhombic transition line closely tracking the paramagnetic-to-antiferromagnetic transition line \cite{Fernandes2014}. While iron-based superconductors are metallic systems, the $J_1$--$J_2$ Heisenberg antiferromagnetic model on the square lattice provides a useful framework to investigate the low-energy properties of the coupled nematic-magnetic degrees of freedom. Besides iron pnictides, the same phenomenology was recently invoked to explain the vestigial nematic phase observed in the heavy-fermion compound CeAuSb$_2$ \cite{Seo2020}. 

    Although critical exponents near the finite-temperature phase transition of the $J_1$--$J_2$ Heisenberg model have shown that the nematic phase transition indeed belongs to the Ising universality class in both classical and quantum cases \cite{Weber2003,Gauthe2022}, evidence from direct phase ordering dynamics has been lacking. When the standard Ising model on a square lattice is quenched from the high-temperature disordered state into the low-temperature ordered state, the growth of the ordered domains exhibits a characteristic power-law behavior, $L(t) \sim t^{1/2}$, known as the Allen-Cahn growth law \cite{Allen1979}. This growth law, a consequence of curvature-driven coarsening, is well-understood from the time-dependent Ginzburg-Landau equation for non-conserved order parameters \cite{Bray1994,Puri2009}. If the Ising-nematic vestigial phase transition truly belongs to the Ising universality class, we expect the dynamics of domain growth to exhibit the same power-law behavior as the standard Ising model in the thermodynamic limit, regardless of the specific details of the dynamics and of the distinct character of the Ising-nematic order parameter. Indeed, in contrast to the standard Ising model, the emergent Ising-nematic order parameter in the $J_1$--$J_2$ Heisenberg model is a composite order formed out of a bilinear combination of the N\'eel vectors, which are the primary order paramaters of the antiferromagnetic transition \cite{Fernandes2019}.

    \begin{figure}[t]
	   \includegraphics[width=1.0\columnwidth]{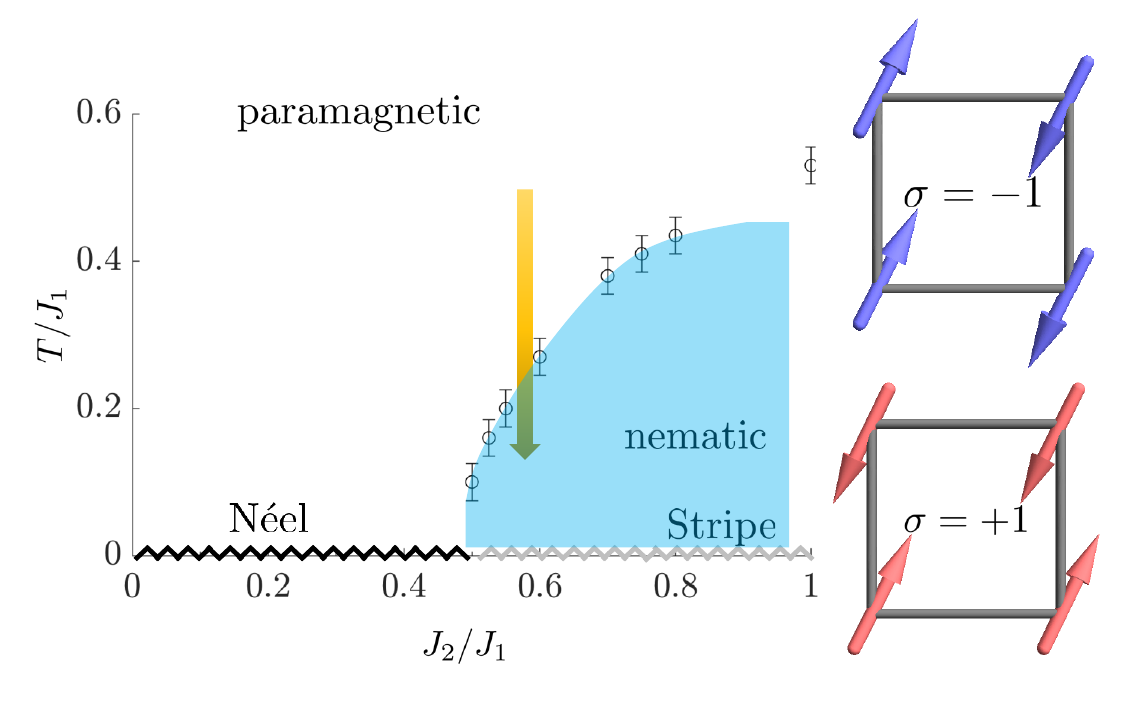}
	   \caption{ 
		Schematic phase diagram of the $J_1$--$J_2$ model and the configuration of the $x$-aligned ($\sigma=+1$) and $y$-aligned ($\sigma=-1$) Ising-nematic order. When $J_2/J_1 > 0.5$, a finite transition temperature $T_c$ exists for the nematic phase. The nonequilibrium dynamics of the system is simulated by quenching the system from $T/J_1=\infty$ to the nematic region.
	   }
	   \label{fig:1} 
    \end{figure}
    
    In this paper, we employ the stochastic Landau-Lifshitz-Gilbert (LLG) equation to examine the relaxation dynamics of the antiferromagnetic $J_1$--$J_2$ Heisenberg model being quenched from $T=\infty$ to $T<T_c$. We find the domain coarsening dynamics is described by two stages. Despite the early nonlinear behavior, $L(t)^2$ asymptotically becomes linear in the late stage, signifying the power-law behavior $L(t)\sim t^{1/2}$. Furthermore, in the presence of random weak bond disorder, we observe a logarithmic domain growth, $L(t)\sim \ln(t)$, which is consistent with the domain growth of the Ising model under weak disorder \cite{Puri1993,Rao1993,Corberi2012}. Different from the kinetic Ising model, which violates the superuniversality hypothesis \cite{Corberi2012}, we show that the equal-time correlation functions at different time collapse onto a single curve when scaled by the characteristic correlation length, suggesting the superuniversality hypothesis is obeyed in our system \cite{Fisher1988,Cugliandolo2010,Corberi2015}. 
    
    The remainder of this paper is structured as follows. Sec.~\ref{sec:method} provides a general description of the $J_1$--$J_2$ Heisenberg model and the stochastic LLG dynamics used for simulating the system's relaxation dynamics. We then reproduce the classical phase diagram of the model to test the validity of our numerical integration scheme and outline our approach for simulating and characterizing the coarsening dynamics of Ising-nematic domains. In Sec.~\ref{sec:results}, we present our numerical simulation results for the correlation functions and the growth of the characteristic domain length. Sec.~\ref{sec:discussion} offers an in-depth discussion on how the two-stage coarsening dynamics influences our measurements of domain growth laws. The effects of weak bond disorder on the coarsening dynamics are investigated in Sec.~\ref{sec:disorder}. Finally, Sec.~\ref{sec:conclusion} concludes the paper by summarizing our key findings and providing an outlook for future research directions, including in connection to materials where vestigial nematicity is realized.

\section{Model and Method}\label{sec:method}
	The classical antiferromagnetic $J_1$--$J_2$ Heisenberg model is given by the Hamiltonian
    \begin{equation}
	   \mathcal{H} = J_1\,\sum_{\langle i,j\rangle }  \mathbf{S}_i \cdot \mathbf{S}_j +J_2\,\sum_{\langle \langle i,j\rangle \rangle } \mathbf{S}_i \cdot \mathbf{S}_j, 
    \end{equation}
    where $\mathbf{S}_i=(S_i^x,S_i^y,S_i^z)$ represents a classical spin degree of freedom with unit length, and $\langle i,j\rangle$ ($\langle\langle i,j \rangle\rangle$) denotes the summation taken over nearest neighbor (next-nearest neighbor) pairs of spins, with the corresponding coupling strength $J_1$ ($J_2$).
    
    The ground states in the strong coupling regime ($J_2/J_1>0.5$) are characterized by two decoupled N\'eel orders residing on the two sublattices of the bipartite square lattice. The two N\'eel vectors can freely rotate relative to each other, corresponding to an $O(3)\times O(3)$ degeneracy. In the presence of fluctuations, this degeneracy is reduced to $O(3)\times \mathbb{Z}_2$ through the order-by-disorder mechanism \cite{Chandra1990}. The $\mathbb{Z}_2$ component corresponds to two nematic states related by a $90^\circ$ rotation, formed by the two possible ways in which the collinear N\'eel vectors can align (parallel or antiparallel). Alternatively, the $\mathbb{Z}_2$ symmetry can be understood as selecting between one of the two magnetic ordering wave vectors $\mathbf{Q}=(\pi,0)$ and $\mathbf{Q}=(0,\pi)$. To spatially capture the nematic order, we take the convention to define a local Ising-type nematic order parameter associated with each square plaquette as \cite{Weber2003}
    \begin{equation}\label{eqn:order_parameter}
    \sigma_{\square} = \frac{(\mathbf{S}_i  - \mathbf{S}_k)\cdot(\mathbf{S}_j  - \mathbf{S}_l)}{\left|(\mathbf{S}_i  - \mathbf{S}_k)\cdot(\mathbf{S}_j  - \mathbf{S}_l)\right|},
    \end{equation}
    where $i$,$j$,$k$,$l$ are the sites arranged counterclockwise on the square plaquette, such that it takes values $+1$ and $-1$, characterizing $x$-aligned stripes and $y$-aligned stripes, respectively.
    
	The classical spin dynamics under the influence of thermal fluctuations at temperature $T$ is simulated by the stochastic Landau-Lifshitz-Gilbert (LLG) equation \cite{Landau1935,Gilbert2004,Garcia-Palacio1998}
    \begin{equation}\label{eqn:LLG}
	\frac{d\mathbf{S}_i}{dt} = \mathbf{S}_i \times ( \mathbf{H}_i + \boldsymbol{\zeta}_i) - \lambda\,\mathbf{S}_i \times [\mathbf{S}_i \times (\mathbf{H}_i+\boldsymbol{\zeta}_i)]
    \end{equation}
    where $\lambda$ is the dimensionless damping coefficient, $\mathbf{H}_i$ denotes the local field from surrounding spins computed by $\mathbf{H}_i= -(\partial \mathcal{H}/\partial S_i^x,\partial \mathcal{H}/\partial S_i^y,\partial \mathcal{H}/\partial S_i^z)$, and $\boldsymbol{\zeta}_i$ is the stochastic field generated by a Gaussian distribution that satisfies
    \begin{align}
        \langle \boldsymbol{\zeta}_i(t) \rangle &= 0,\nonumber\\
        \langle \zeta^\alpha_i(t)   \zeta^\beta_j(t')\rangle &= 2\lambda k_B T \delta_{i,j}\delta_{\alpha,\beta}\delta(t-t'),
    \end{align}
    with $\alpha,\beta =x,y,z$. The gyromagnetic ratio $\gamma$ is absorbed together with $J_1$ and the spin length $S$ into the time $t$, such that $t=1$ corresponds to the typical characteristic time scale $\tau=(\gamma J_1S)^{-1}$ in picoseconds for spin dynamics.

    \begin{figure}[t]
	\includegraphics[width=1.0\columnwidth]{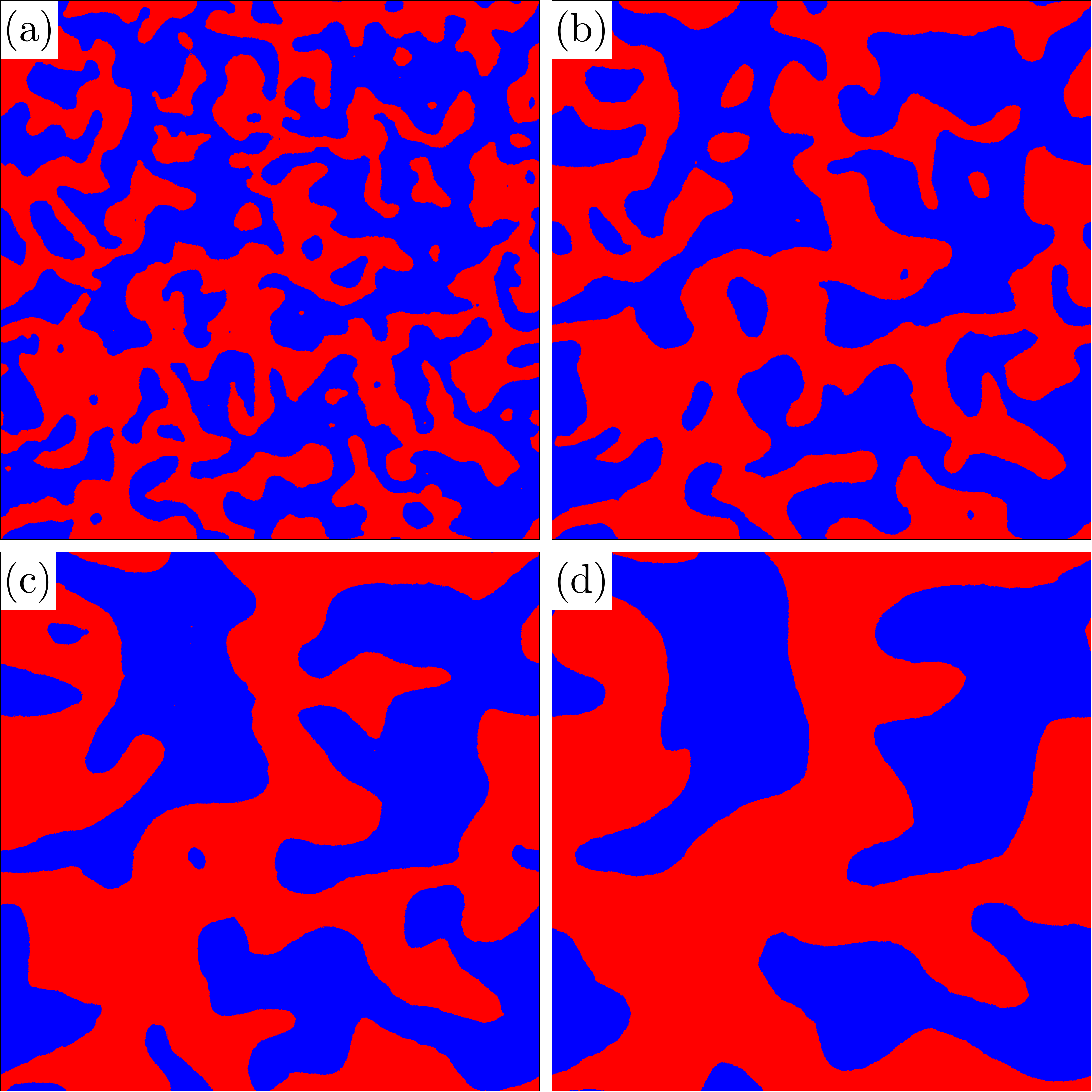}
	\caption{ 
		Snapshots of the coarsening of the Ising-nematic domains with $J_2/J_1 = 0.55$ and $T/J_1 = 0.01$ on a $2048\times 2048$ square lattice at (a) $t = 250$, (b) $t = 500$, (c) $t = 1000$, and (d) $t = 2000$. The red and blue regions correspond to Ising-nematic domains with the local order parameter $\sigma_\square=+1$ and $-1$, respectively.
		%The time steps interval in this simulation is $\delta t = 0.05$ and the damping coefficient is $\lambda = 0.05$.
    }				
	\label{fig:2}
    \end{figure}
    
    The numerical integration scheme of the stochastic LLG equation (\ref{eqn:LLG}) is adapted from a generic 4th order Runge-Kutta (RK4) method for integrating Stratonovich stochastic differential equations \cite{Gard1988}. The validity of our numerical integration scheme is checked by reproducing a schematic classical phase diagram of the $J_1$--$J_2$ model, where the critical temperature $T_c$ is obtained by locating the peak of the specific heat averaged over $100$ simulation runs on a small cluster of the size $128\times 128$ (shown in Fig.~\ref{fig:1}). When $J_2/J_1 < 1/2$, the system orders into the N\'eel state at $T/J_1=0$, characterized by the wave vector $\mathbf{Q}=(\pi,\pi)$. At finite temperatures, the order is immediately destroyed by thermal fluctuations, leaving the system in the paramagnetic state. When $J_2/J_1>1/2$, the system orders into one of the two stripe magnetic states with wave vectors $\mathbf{Q}=(\pi,0)$ and $\mathbf{Q}=(0,\pi)$. At finite temperatures, nematic order ($\sigma_\square=+ 1$ or $-1$) onsets below a critical temperature as a function of the ratio $J_2/J_1$, consistent with previous results obtained from Monte Carlo simulations \cite{Weber2003}.
    
    Having reproduced the phase diagram, we proceed to investigate how the system develops nematic order when the system is quenched from $T/J_1=\infty$ to a temperature below the critical point. At $t=0$, we initialize the system on a $2048\times 2048$ square lattice with random spin orientations, representing the high-temperature disordered state. Each simulation runs to $t=2000$ with a time step $\delta t=0.05$ and a damping factor $\lambda=0.05$ \footnote{The effects of varying $\lambda$ are discussed in Sec.\ref{sec:discussion} more specifically. Since varying $\lambda$ will not change our main results, we fix $\lambda=0.05$ in all our simulations.}. 
    Fig.~\ref{fig:2} shows snapshots of the coarsening of the Ising-nematic domains at $J_2/J_1 = 0.55$ and $T/J_1 = 0.01$, characterized by the local nematic order parameter defined in Eq.~(\ref{eqn:order_parameter}). For $\sigma_\square = +1$ (red), the domain is associated with $x$-aligned stripes while for $\sigma_\square = -1$ (blue) the domain is associated with $y$-aligned stripes. We emphasize that there is no long-range magnetic order, and that the Ising-nematic order corresponds to the locking between the relative orientation between the two N\'eel vectors. 

    We characterize the growth of the nematic domain by the equal-time correlation function 
    \begin{equation}
	   C(r,t)\equiv C(\mathbf{r},t) = \langle \sigma(\bs{r}_0,t) \sigma(\bs{r}_0 + \bs{r},t) \rangle,
    \end{equation}
    where the bracket $\langle ...\rangle $ here denotes averaging over all different independent initializations and all reference sites $\mathbf{r}_0$ in the system. The equal-time correlation function measures the spatial correlation between the local nematic order parameter $\sigma_\square$ at positions $\mathbf{r}_0$ and $\mathbf{r}_0 + \mathbf{r}$ at time $t$. In this work, we average over $100$ independent simulation runs to minimize statistical errors in the calculated correlation function. 
    
    From the equal-time correlation function, we can extract the characteristic domain length $L(t)$ by exploiting the scaling property of the correlation function
    \begin{align}
        C(r,t)=f(\xi),
    \end{align}
    where we define the rescaled length $\xi\equiv r/L(t)$ up to a constant ratio such that $f(\xi)$ becomes a universal scaling function independent of time and other model parameters, indicating the dynamic scaling invariance of the phase ordering process \cite{Bray1994, Puri2009}. A convenient choice for determining the characteristic domain length $L(t)$ is given by the condition $C(L(t),t)=C(0,t)/2$, which states that the correlation function is reduced by half at $L(t)$ from its self correlation. This analysis typically reveals a power-law growth of the characteristic length $L(t)\sim t^{n}$ if the system exhibits dynamic scaling invariance. By determining the value of the exponent $n$, we can also identify the universality class of the phase ordering dynamics.

\section{Coarsening dynamics of nematic domains}\label{sec:results}
\begin{figure}[h]
	   \includegraphics[width=\linewidth]{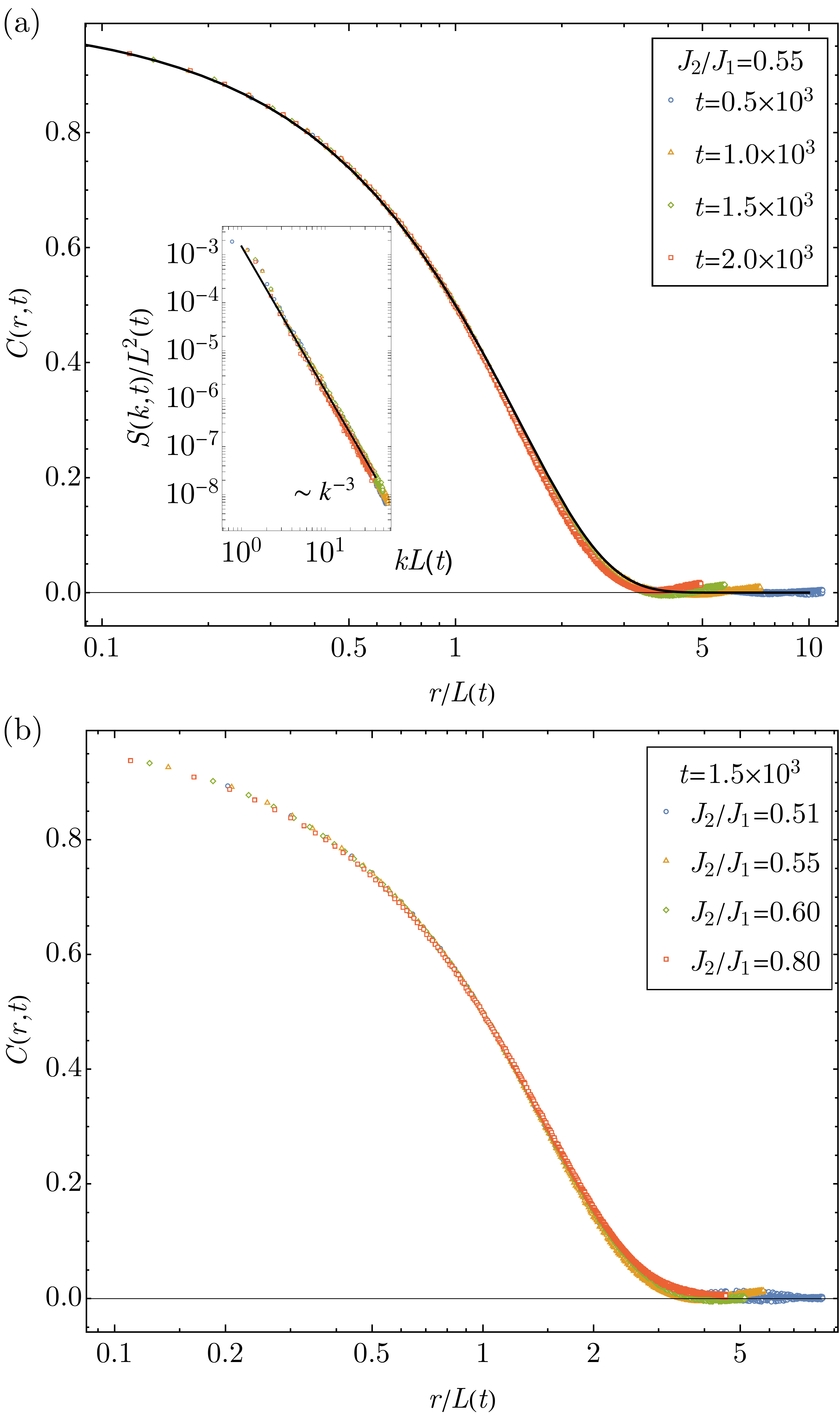}
	   \caption{(a) Dynamic scaling for the equal-time correlation function $C(r,t)$ for $J_2/J_1=0.55$ at $T/J_1=0.01$ in the linear--log scale for various times. The collapsed $C(r,t)$ is fitted with the OJK form of the correlation function $f_{\mathrm{OJK}}(\xi)$ (black solid line).  The inset shows that the structure factor tail follows the Porod's law $S(k)\sim k^{-3}$ in the log--log scale (additional plots for $C(r,t)$ with various $J_2/J_1$ can be found in the App.~\ref{Sec:A1}). (b) Dynamic scaling for the equal-time correlation function $C(r,t)$ in the linear--log scale for various ratios of $J_2/J_1$, showing the universal behavior of $C(r,t)$ independent of the ratio between $J_1$ and $J_2$ within the nematic regime.
	   }
	   \label{fig:3}
    \end{figure}

    We first examine the dynamic scaling invariance of the correlation function at various times and various coupling constants $J_2/J_1$. Fig.~\ref{fig:3} (a) and Fig.~\ref{fig:3} (b) show the collapse of the correlation function $C(r,t)$ onto a single curve after rescaling $r/L(t)$ for various times and various coupling constants, respectively. The rescaled correlation function can be fitted with the Ohta-Jasnow-Kawasaki (OJK) form of the correlation function \cite{Ohta1982}
    \begin{align}\label{eqn:OJK}
        f_{\mathrm{OJK}}(\xi)\equiv\frac{2}{\pi}\sin^{-1}\left[\exp(-\xi^2)\right],
    \end{align}
    which also describes the correlation function for the kinetic Ising model \cite{Bray1994,Puri2009}. The collapse of all the correlation function $C(r,t)$ after the rescaling indicates a universal behavior of the correlation function that only depends on the characteristic domain length $L(t)$.
    
    Moreover, the short-distance behavior of $f(\xi)$ gives rise to a power-law decay behavior near the tail of the structure factor
    \begin{align}
        S(k)/L^d\sim (kL)^{d+1}
    \end{align}
    where $d$ stands for the dimension of the system, and the structure factor $S(k)$ is computed directly from the Fourier transform of the correlation function. This scaling behavior of the power-law tail is referred as the Porod's law \cite{Porod1982}. As it is shown in the inset of Fig.~\ref{fig:3} (a), the structure factor can be fit nicely with a power-law tail $\sim k^{-3}$ (with $d=2$), which is in agreement with the power-law behavior in the kinetic Ising model \cite{Bray1994}.

\begin{figure}[t]
	   \includegraphics[width=1.0\columnwidth]{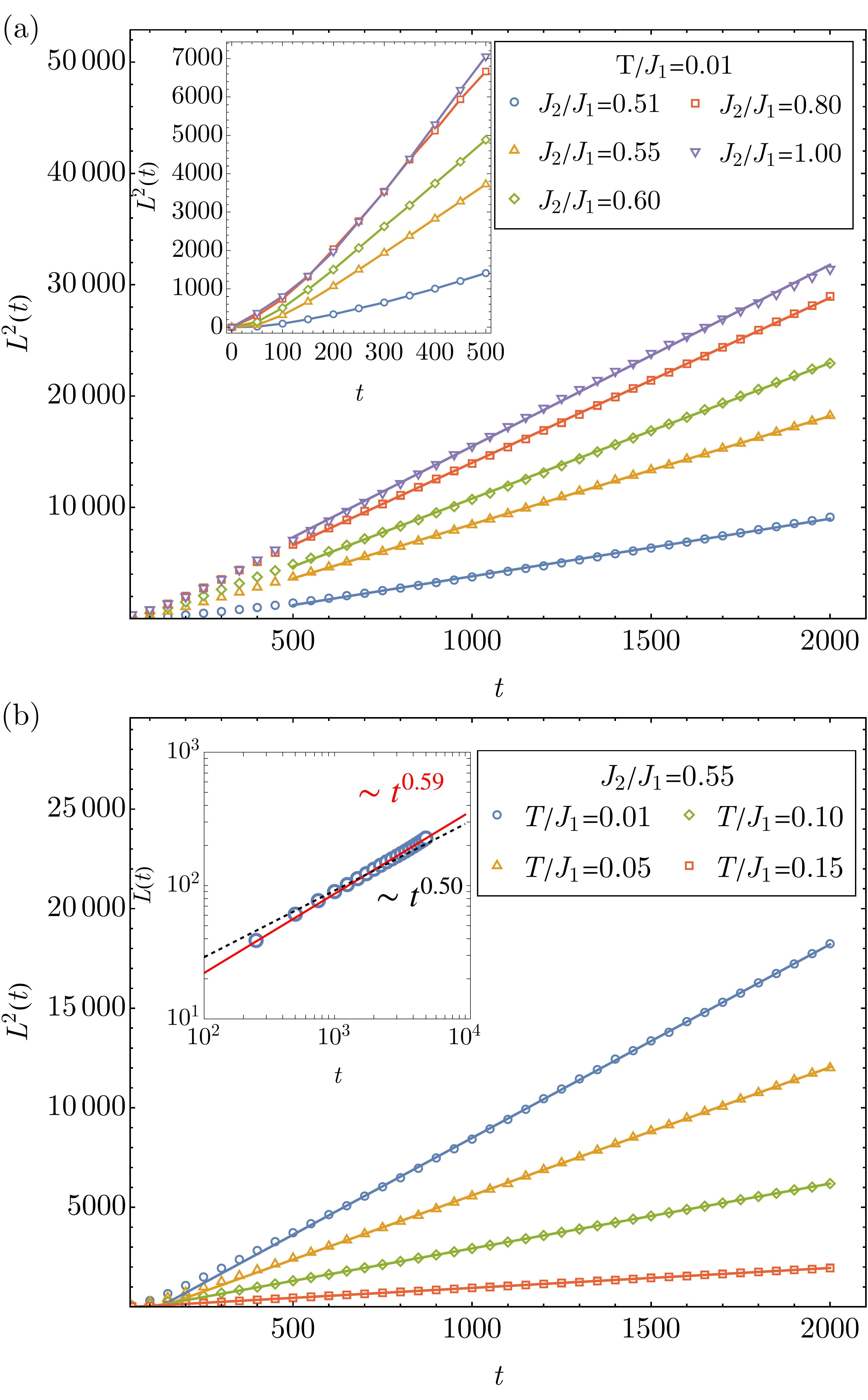}
	   \caption{(a) Time evolution of the squared characteristic domain length $L^2(t)$ at various $J_2/J_1$ with the fixed temperature $T/J_1=0.01$. The inset shows that the early-stage dynamics behaves differently from the asymptotic dynamics at later time. (b) Time Evolution of the characteristic domain area $L^2(t)$ at various $T/J_1$ with the fixed constant coupling $J_2/J_1=0.55$. The inset (log--log scale) shows that the fitting $L(t) \sim t^n$ deviates from the Allen-Cahn's law.}
	   \label{fig:4}  \
    \end{figure}

    Fig.~\ref{fig:4} (a) and (b) present the time evolution of the characteristic domain length by plotting the square of the characteristic length, $L^2(t)$, as a function of time $t$. In 2D, $L^2(t)$ represents the characteristic area $A$ of a nematic domain. Using the generalized Gauss-Bonnet theorem in 2D \cite{Sicilia2007}, we can obtain the changing rate of a simply-connected domain in the curvature-driven coarsening process as
    \begin{align}\label{eqn:area}
        \frac{dA}{dt}\sim \oint_{\partial A} \kappa\,\mathrm{d}l=\pm 2\pi,
    \end{align}
    where $\kappa$ denotes the (signed) curvature of the domain boundary \footnote{The signed curvature is positive when following a curve counterclockwise and negative when following it clockwise, with the magnitude remaining the same in both directions.}, and the integration is performed over the boundary $\partial A$ of the domain. Consequently, we have $A \sim t$, which implies that $L^2(t)$ should also have a linear time dependence, regardless of the shape and size of the domain. We then arrive at the Allen-Cahn growth law $L(t) \sim t^{1/2}$. As shown in Fig.~\ref{fig:4} (a) and (b), $L^2(t)$ indeed exhibits a linear behavior in the late stage ($t > 1000$) of the coarsening process for various temperatures and various ratios of $J_2/J_1$ within the nematic regime. This indicates the late-stage coarsening dynamics satisfies the Allen-Cahn law.
    
    In contrast, the early stage ($t < 500$) of $L^2(t)$ is characterized by a distinctly nonlinear time evolution of $L^2(t)$, as highlighted in the inset of Fig.~\ref{fig:4} (a). A direct consequence of this nonlinear behavior is that we cannot recover the Allen-Cahn law when plotting $L(t)$ against $t$ and trying to fit it with a single power law. Even when extending the simulation to $t = 5000$ and fitting the power-law $L(t) \sim t^n$ only in the late stage of the evolution ($t>1000$),  we still observe a significant deviation from the expected exponent of $1/2$ predicted by the Allen-Cahn law, as shown in the inset of Fig.~\ref{fig:4} (b)]. This apparent contradiction with the results obtained from fitting $L^2(t)$ with a linear function raises questions about whether the coarsening dynamics of nematic domains truly belongs to the Ising universality class.

\section{Two-stage coarsening dynamics}\label{sec:discussion}
\begin{figure}[t]
	   \includegraphics[width=0.8\columnwidth]{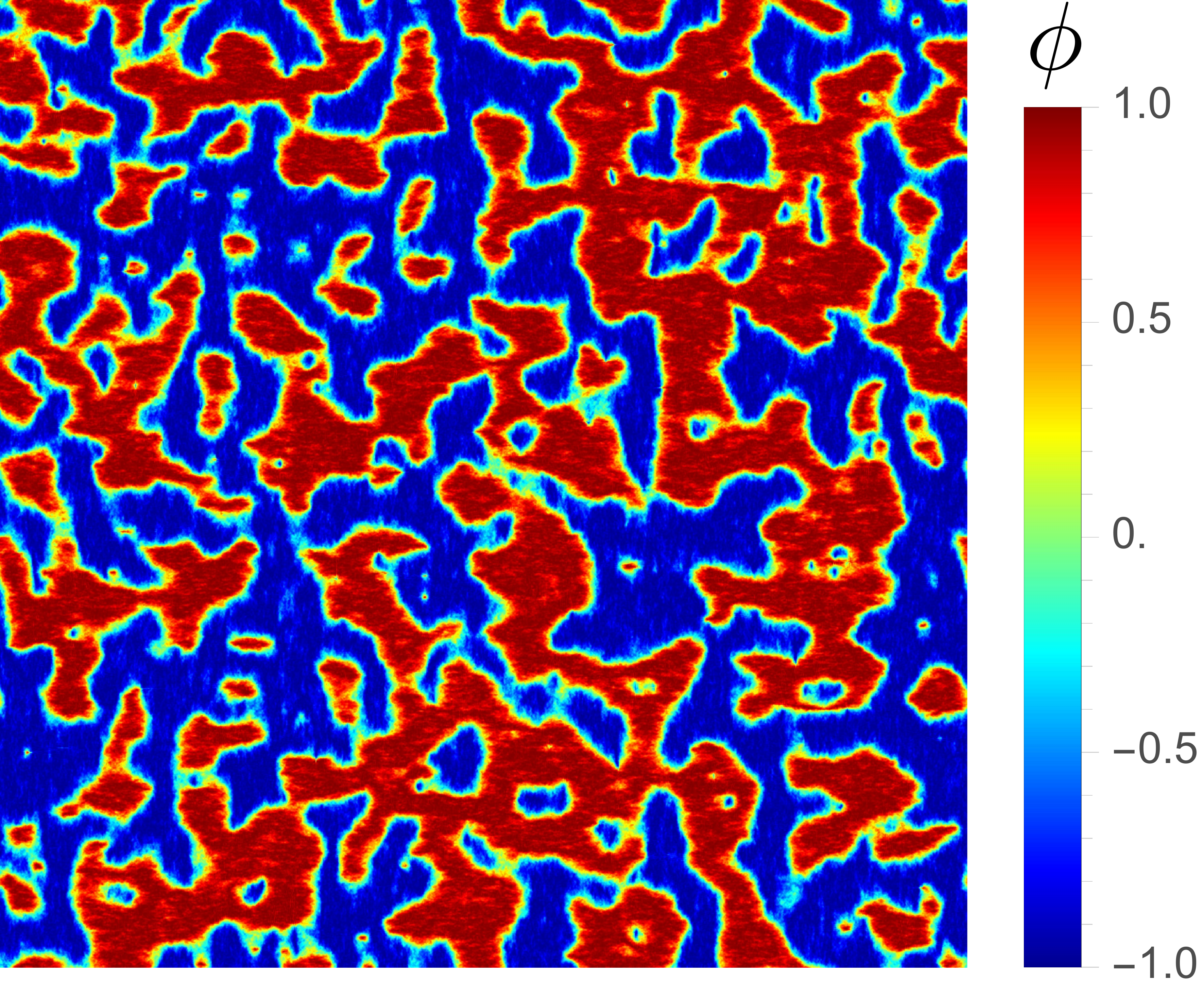}
	   \caption{Early-stage $(t=100)$ nematic domains depicted by the scalar local order parameter $\phi_\square$ on a $1024\times 1024 $ lattice with $J_2/J_1=0.55$ and $T/J_1=0.01$.}
	   \label{fig:5}
    \end{figure}

   To understand the origin of this discrepancy, it is essential to consider the assumptions underlying the Allen-Cahn law. The Allen-Cahn law, which describes a curvature-driven coarsening process, implies that the domains in the system have smooth boundaries with fully saturated local order parameters inside. However, when our system is initialized with a random configuration of spins at $t = 0$, local order parameters are also randomized being far from their saturated values in thermal equilibrium. Moreover, the system does not have smooth boundaries for the domains formed at this early stage. This information is partially obscured when we measure the nematic domains using the Ising-type order parameter $\sigma_\square$ because $\sigma_\square$ only takes values of $\pm 1$, making it impossible to observe the coarsening towards saturation of the order parameter. If we define a scalar local nematic order parameter from Eq.~(\ref{eqn:order_parameter}) as
    \begin{align}\label{eqn:order_parameter_scalar}
        \phi_\square = \frac{1}{4} (\mathbf{S}_i  - \mathbf{S}_k)\cdot(\mathbf{S}_j  - \mathbf{S}_l),
    \end{align}
    we can see that, apart from a few large domains, the system consists of many small irregular domains that are not fully saturated (shown in Fig.~\ref{fig:5}, more comparison between saturated domains and unsaturated domains can be found in App.~\ref{Sec:A3}). The coarsening of these small domains cannot be described by the curvature-driven dynamics. Hence a non-Allen-Cahn behavior dominates the early-stage dynamics. When we consider the growth of the characteristic domain size in the late stage, the early-stage dynamics introduces an offset time $t_0$, which corresponds to the onset time of the curvature-driven coarsening such that 
    \begin{align}
        L_{\mathrm{late}}(t) \sim \sqrt{t-t_0}.
    \end{align}
    
\begin{figure}[t]
        \includegraphics[width=\columnwidth]{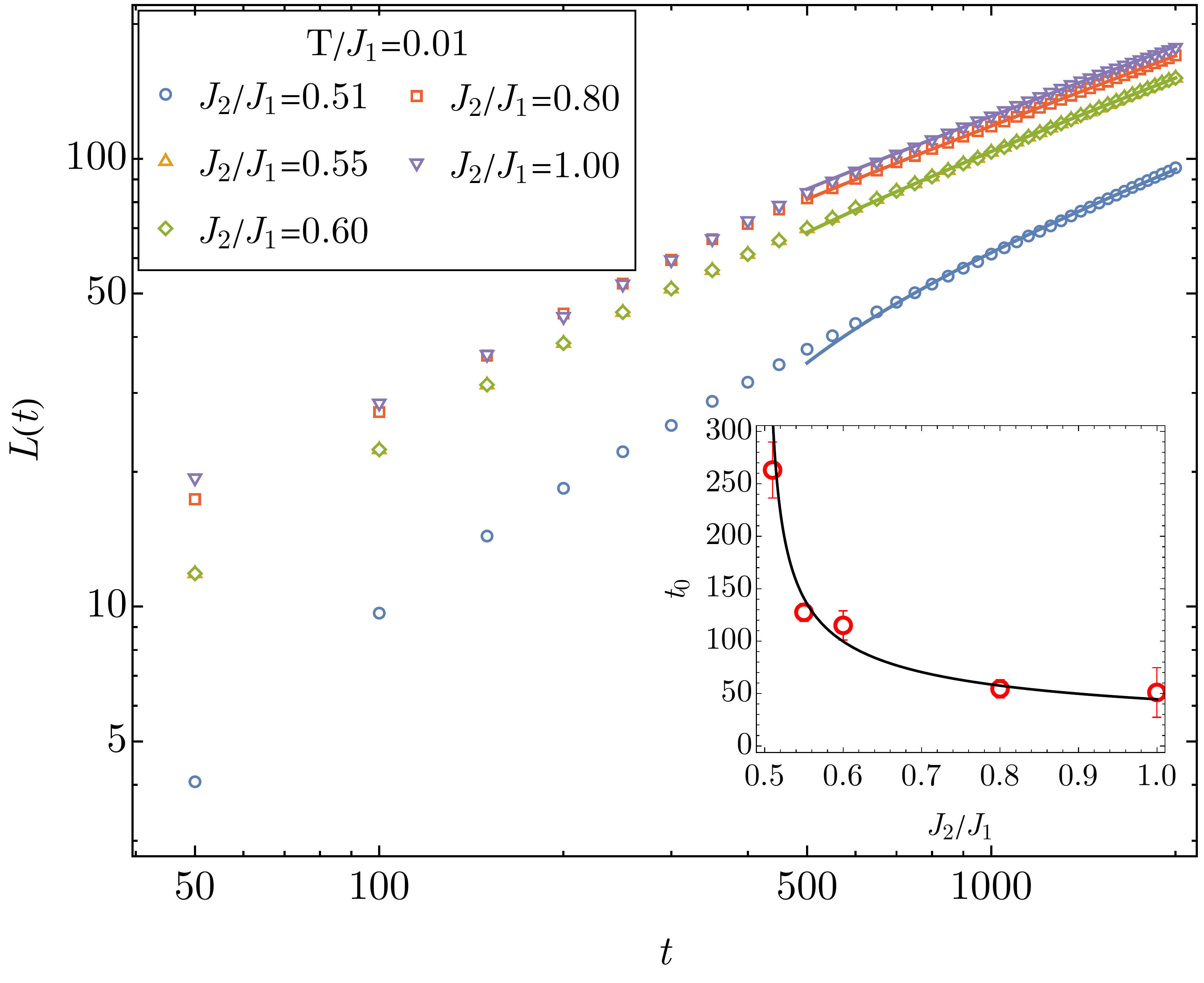}
        \caption{Time evolution of the characteristic domain length $L(t)$ in the log--log scale fitted with $\sqrt{t-t_0}$. The $t_0$ obtained from the fitting is shown in the inset, which behaves like $t_0\sim |J_2/J_1-1/2|^{-1/2}$ (black curve).}
	   \label{fig:6}
    \end{figure}

    Therefore, when $t_0$ is sizable, an accurate measurement of the asymptotic Allen-Cahn $1/2$ growth law requires us to examine the coarsening dynamics at time scales several orders of magnitude longer than our current simulation time, in a system that is also several orders of magnitude larger than our current setup. This is beyond our computational capabilities. However, as we move deeper into the nematic phase by increasing the ratio of $J_2/J_1$, the local nematic order parameter is saturated more quickly, such that the effect of the onset time $t_0$ becomes less significant.  This is shown in Fig.~\ref{fig:6}. As we fit $L(t)$ with the power law $\sim \sqrt{t-t_0}$ for $t > 1000$ for the systems with larger $J_2/J_1$ ratios, $t_0$ becomes smaller and smaller. On the other hand, when we directly examine the growth of the area $L^2(t)$, the onset time for the curvature-driven coarsening is already incorporated into the linear fitting, as it simply shifts the curve. Hence, we can easily deduce the asymptotic Allen-Cahn power-law behavior for $L(t)$ from $L^2(t)$. 
    
\begin{figure}[t]
	   \centering
        \includegraphics[width=\columnwidth]{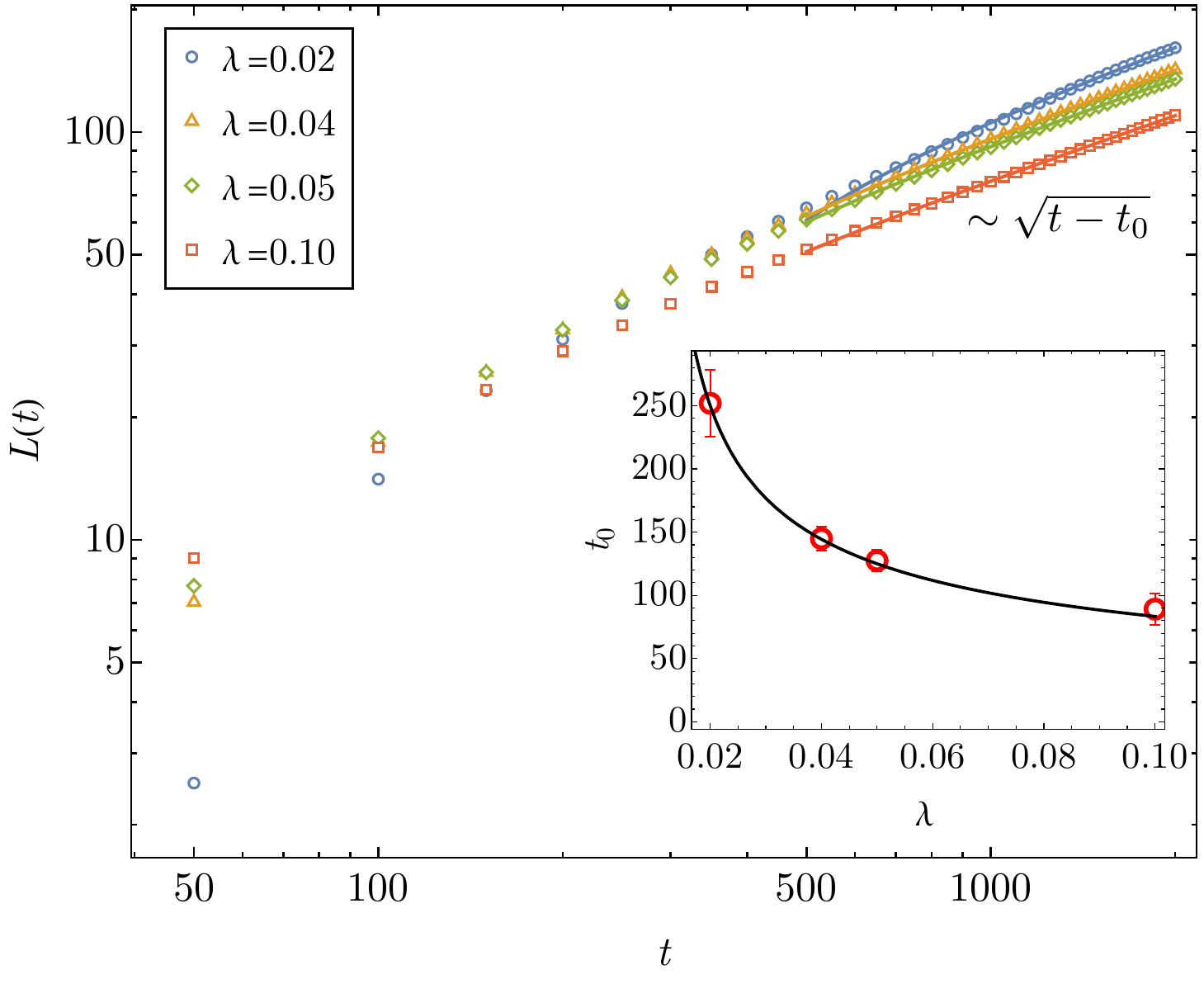}
	   \caption{Time evolution of the characteristic domain length under various damping coefficient $\lambda$ at $J_2/J_1=0.55$ and $T/J_1=0.01$. The inset shows $t_0$ decreases with increasing $\lambda$.
    }\label{fig:7}
    \end{figure}

    Furthermore, the dependence of characteristic domain length on damping constant $\lambda$ in the stochastic LLG equation (\ref{eqn:LLG}), shown in Fig.~\ref{fig:7}, provides further evidence for two-stage coarsening dynamics. Larger $\lambda$ produces smaller offset time $t_0$ but slower overall domain growth. This dichotomy reflects the distinct mechanisms governing each stage. According to the fluctuation-dissipation theorem, the larger damping coefficient gives a larger stochastic field $\boldsymbol{\zeta}_i$, accelerating the process to reach local equilibrium and reducing $t_0$. Conversely, in the late stage, where domain wall curvature drives the dynamics, these stochastic fields disrupt the deterministic curvature-driven motion, effectively slowing domain growth. It is noteworthy that a similar multi-stage coarsening has also been observed in the kinetic Ising model \cite{Fialkowski2002}. However, for the kinetic Ising model in 2D, the early-stage dynamics gives rise to a similar $1/2$ power-law behavior as the late-stage dynamics. Consequently, the power-law fitting for the late stage is not affected by the early-stage dynamics in the kinetic Ising model. This strongly suggests that the early-stage coarsening dynamics is model-dependent, as evidenced by another example where LLG dynamics of a different Hamiltonian shows an $L(t)\sim t^{1/3}$ coarsening behavior in the early stage \cite{Kudo2016}.

\begin{figure}[t]
	   \includegraphics[width=1.0\columnwidth]{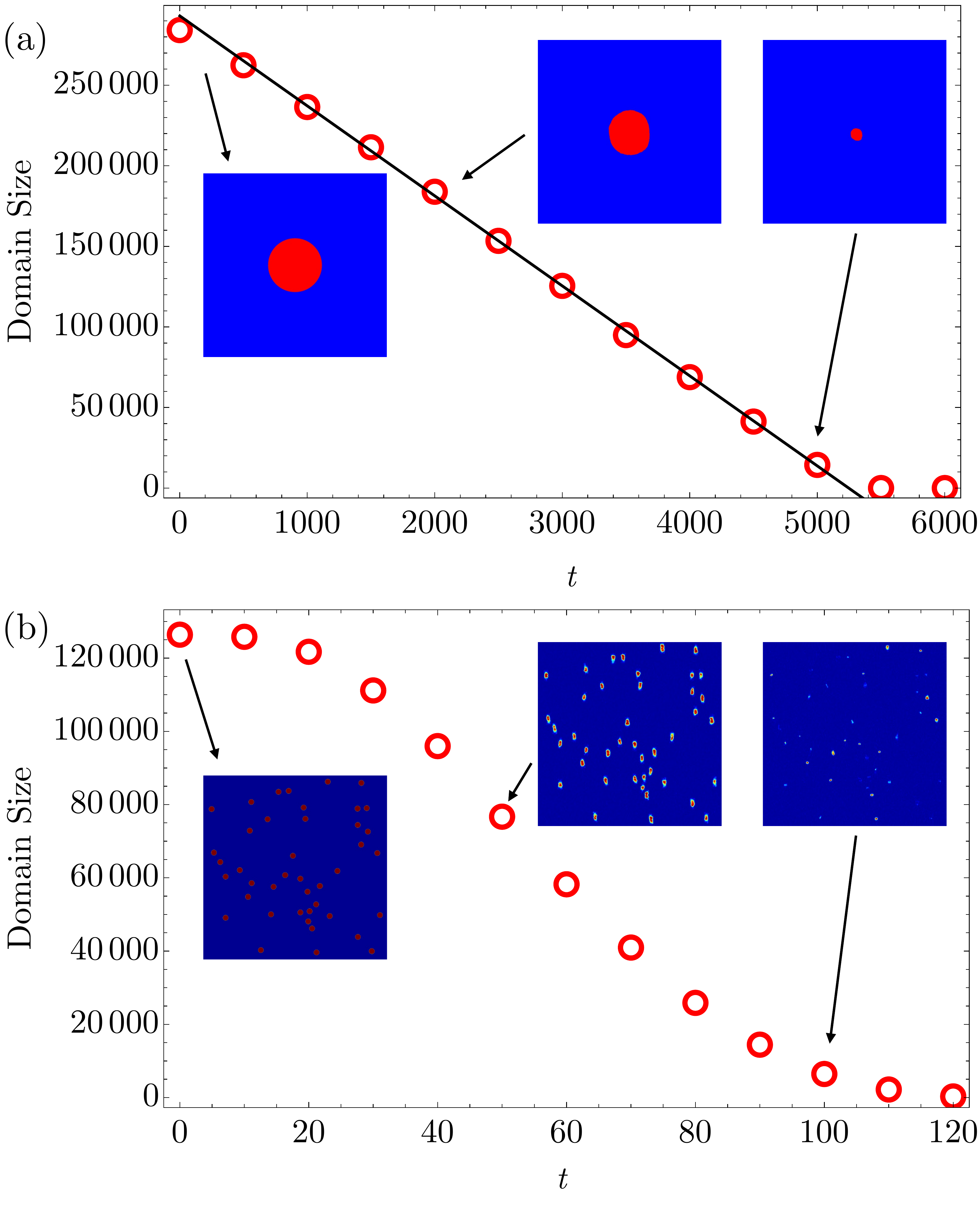}
	   \caption{(a) Coarsening of a single large disk domain with an initial radius $R_0=300$ embedded in a $2048\times 2048$ system at $J_2/J_1=0.55$ and $T/J_1=0.01$. The snapshots depict the normalized Ising-nematic order parameter $\sigma_\square$. (b) Coarsening of non-touching small disk domains with $R_0=30$ randomly scattered across the system under the same model parameters. The snapshots depict the non-normalized scalar order parameter $\phi_\square$. The domain size is computed by summing over all small domains}
	\label{fig:8} 
    \end{figure}

    Finally, we demonstrate that the late-stage coarsening dynamics is indeed curvature-driven and distinct from the early-stage coarsening dynamics by considering two coarsening scenarios. In the first scenario, we prepare an initial configuration consisting of two nematic domains. At $t = 0$, we have one disk-shaped domain with a radius $R_0 = 300$ lattice spacings, fully saturated in the $\sigma_\square = +1$ nematic order, surrounded by a fully saturated $\sigma_\square = -1$ domain on a $2048\times 2048$ square lattice. We set $J_2/J_1 = 0.55$ and $T/J_1 = 0.01$, and allow the system to evolve from this artificial configuration. The results are shown in Fig.~\ref{fig:8} (a). We measure the size of the $\sigma_\square = +1$ domain and obtain a linear decrease in the domain size for the circular disk domain, in agreement with Eq.~(\ref{eqn:area}), as predicted by curvature-driven coarsening. In the second scenario, we prepare small disk domains with $R_0 = 30$ scattered throughout the same system to mimic the small domains present in the early stage of the coarsening process. We allow the system to evolve from this configuration under the same parameters as in our first scenario. The results are presented in Fig.~\ref{fig:8} (b). We obtain a nonlinear coarsening behavior for the domain size, similar to the early-stage coarsening shown in the inset of Figure \ref{fig:4}(a). From the snapshots of the coarsening process, depicted using the non-normalized scalar local order parameter $\phi_\square$, we can also see that the accelerated coarsening in the early stage is associated with the saturation of the local order parameter (from dark domains to bright domains). By comparing the two scenarios, we can draw clear distinctions between the early-stage and late-stage coarsening dynamics, and further confirm the $1/2$ asymptotic power law for $L(t)$.
    
\section{Coarsening under weak bond disorder}\label{sec:disorder}
    \begin{figure}[t]
	   \includegraphics[width=0.99\columnwidth]{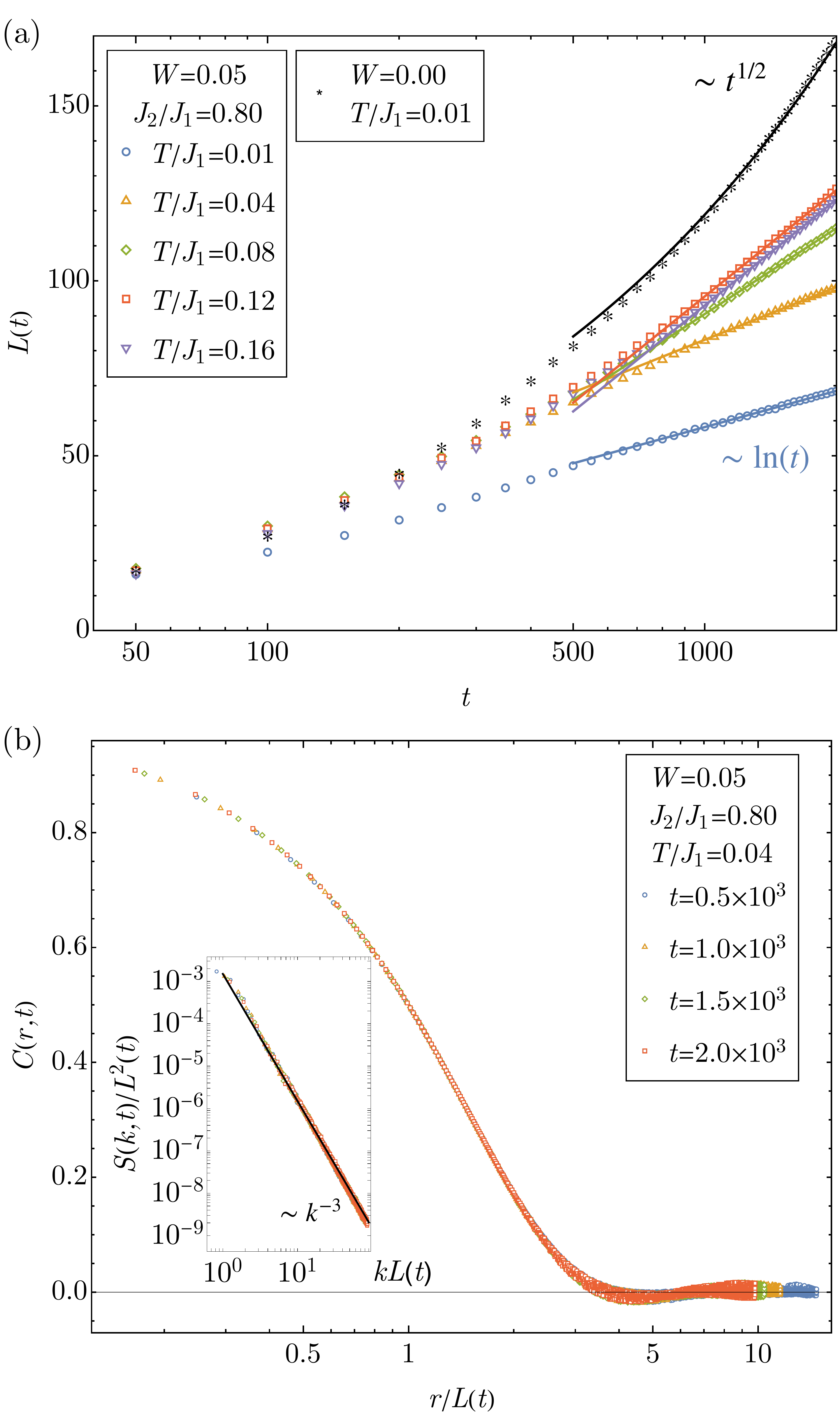}
	   \caption{(a) Time evolution of the characteristic domain length $L(t)$ for $W=0.05$, $J_2/J_1=0.80$ at various temperatures in the linear--log scale. The asymptotic behavior of $L(t)$ approaches $\sim \ln(t)$, deviating from the disorder-free case $L(t)\sim t^{1/2}$. (b) Dynamic scaling of $C(r,t)$ at $T/J_1=0.04$ in the linear--log scale. The inset shows that the structure factor tail satisfies Porod's law $S(k)\sim k^{-3}$ in the log--log scale (Additional plots for $C(r,t)$ under other weak disorder strength $W$ can be found in the App.~\ref{Sec:A2})}.
	   \label{fig:9} 
    \end{figure} 
    In addition, we investigate the effect of weak bond disorder, $J_1 + \delta J_1$, with $\delta J_1/J_1 \in [-W/2, W/2]$, on the coarsening of the nematic domain \cite{Miranda2021}. The weak random disorder on $J_1$ effectively generates a random field for the local Ising nematic order parameter, which prevents our 2D system from developing true long-range nematic order according to the well-known Imry-Ma argument \cite{Imry1975}. In this case, the system always breaks up in nematic domains beyond a characteristic breakup length that decreases with increasing disorder strength as $l_b \sim \exp\left[ \left( J_1/W \right) ^2\right]$ \cite{Meese2022}. Particularly when the disorder is weak, the breakup length $l_b$ can be significantly larger than the lattice scale of our simulations. Therefore, it is still meaningful to study the coarsening dynamics of these nematic domains before their sizes reach the breakup length. 
    
    In our study, to minimize the influence of the early-stage coarsening dynamics on the late stage, we set $J_2/J_1 = 0.8$ and $T/J_1 = 0.01$, ensuring that the system is deep in the nematic regime while keeping all other simulation parameters unchanged. We first set $W = 0.05$ and allow the system to evolve under different temperatures. The results are shown in Fig.~\ref{fig:9}(a) and (b). Although the weak bond disorder suppresses the time evolution of the characteristic length $L(t)$ from power-law growth to logarithmic growth, the correlation functions still collapse onto a single curve, and their Fourier transforms satisfy Porod's law. These observations are consistent with the superuniversality hypothesis, which states that the effect of weak disorder can be fully accounted for by the suppression of the characteristic domain length growth \cite{Fisher1988,Cugliandolo2010,Corberi2015}. Furthermore, the logarithmic growth of $L(t)$ agrees with the general $\sim (\ln t)^{1/\varphi}$ behavior observed in the kinetic Ising model under weak random-field disorder \cite{Puri1993,Rao1993,Corberi2012}.
    
\begin{figure}[t]
	   \includegraphics[width=0.99\columnwidth]{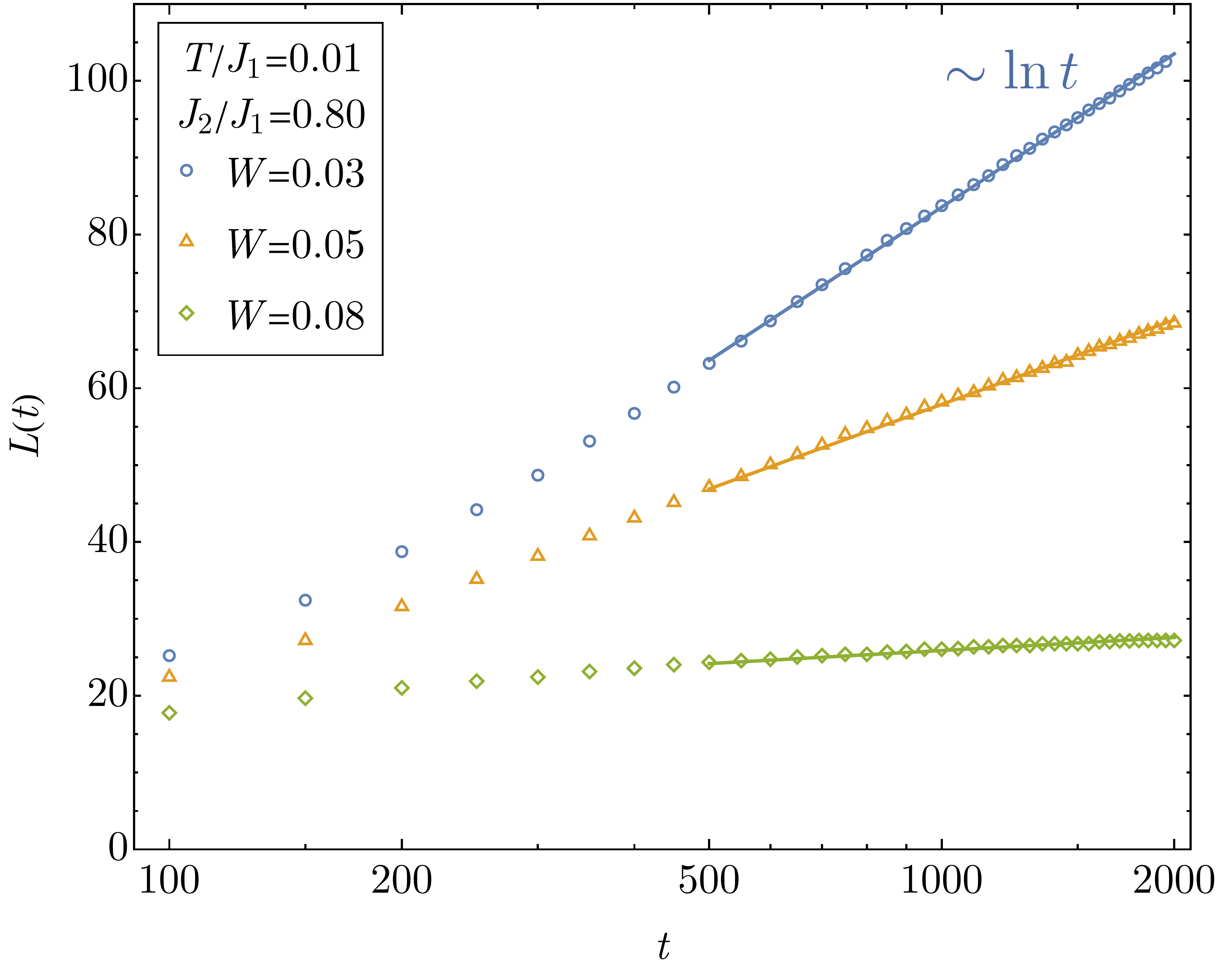}
	   \caption{(a) Time evolution of the characteristic domain length scale $L(t)$ for various $W$ at $T/J_1=0.01$ in the linear--log scale. The asymptotic behavior of $L(t)$ approaches $\sim \ln(t)$.}
	   \label{fig:10} 
    \end{figure}

    However, in the kinetic Ising model, the power $1/\varphi$ varies with respect to different disorder strengths, instead of being fixed at $1/\varphi = 1$. More importantly, a crossover behavior of $C(r,t)$ is observed in the kinetic Ising model, leading to the violation of the superuniversality hypothesis \cite{Corberi2012}. This crossover behavior is absent in our system. We measure the growth of $L(t)$ under different strengths of the weak bond disorder $W$, as shown in Fig.~\ref{fig:10}. The asymptotic behaviors of $L(t)$ for various $W$ approach $\sim \ln t$, showing no dependence of the disorder strength on $1/\psi$. These qualitative differences between our system and the kinetic Ising model suggest that the coarsening dynamics in the presence of random disorder may be affected by the details of the microscopic models. 

\section{Conclusion}\label{sec:conclusion}
    
    In this work, we have investigated the phase ordering dynamics of the classical antiferromagnetic $J_1$--$J_2$ Heisenberg model on a square lattice in the strong frustration regime ($J_2/J_1 > 1/2$) using the stochastic Landau-Lifshitz-Gilbert (LLG) equation. By analyzing the equal-time correlation function and the growth of the characteristic domain length, we have shown that the system undergoes a two-stage coarsening process after being quenched from a random initial state into the nematic phase.

    In the early stage, the local nematic order parameter evolves from purely random initialization, leading to the formation of small, non-saturated domains. The coarsening dynamics of these domains is characterized by a distinctly nonlinear evolution of the domain area characterized by $L^2(t)$, which cannot be described by the curvature-driven Allen-Cahn law. This nonlinear behavior may have a significant impact on the direct measurement of the growth exponent when fitting the characteristic domain length $L(t)$ to a power law, even when considering only the late stage of the evolution. 
    
    The late stage of the coarsening process is dominated by the growth of larger domains with smooth boundaries, whose characteristic areas $L^2(t)$ exhibit a linear time dependence. This behavior is consistent with the curvature-driven coarsening dynamics described by the Allen-Cahn law, which predicts a growth exponent of $1/2$ for the characteristic domain length. It also provides numerical evidence that the Ising-nematic transition in the antiferromagnetic $J_1$--$J_2$ Heisenberg model belongs to the Ising universality class also in what concerns the phase ordering dynamics. 

    Under weak bond disorder, we observe that domain growth is suppressed from power-law to logarithmic behavior. Notably, the system maintains dynamic scaling invariance, with correlation functions collapsing onto a single curve and following Porod's law, consistent with the superuniversality hypothesis. In contrast, the kinetic Ising model violates the superuniversality hypothesis. This qualitative discrepancy suggests that coarsening dynamics under weak disorder may be sensitive to the details of the microscopic models, but this does not rule out the possibility that our simulations have yet to reach the true asymptotic regime. This question remains open for future investigation.
    
    A vestigial Ising-nematic phase, similar to the one emerging in the $J_1$--$J_2$ model, is believed to be realized in the iron pnictides \cite{Fernandes2014} and in the heavy-fermion compound CeAuSb$_2$ \cite{Seo2020}. Although strong electron correlation in these compounds is inherently quantum in nature, the underlying spin dynamics behind the dynamics of the nematic order parameter field can still be treated semi-classically through adiabatic approximation, assuming the order parameter evolves much slower than the electronic degrees of freedom. This approach has been shown to provide very good agreement with exact quantum dynamics in the Hubbard model with emergent spin density waves \cite{Chern2018}. Quantum effects for spins with $S\neq 1/2$ can addressed in a similar semi-classical approach by studying the Langevin dynamics of generalized spins with coherent $SU(N)$ states \cite{Dahlbom2022}. 
    
    Therefore, in view of our theoretical results, it would be interesting to experimentally probe the nematic domain dynamics in these systems. While directly observing nematic domain coarsening remains challenging, several complementary techniques offer promising approaches. Time-resolved resonant X-ray scattering can detect transient electronic anisotropy associated with nematic ordering \cite{Liu2018}; ultrafast optical pump-probe spectroscopy \cite{Patz2014} detects nematic fluctuations and reveals strong magnetoelastic coupling; ultrafast electron diffraction \cite{Konstantinova2019} tracks structural distortions from nematic response to laser pulse perturbation; and time-resolved electron microscopy \cite{Flannigan2022} provides spatial mapping for the dynamics of the lattice under nematic fluctuations. These techniques may reveal our predicted scaling behaviors through analysis of relaxation times and characteristic length scales. Ref. ~\cite{Klein2020} also put forward a theoretical proposal to control nematic order out-of-equilibrium via laser-induced phonon excitations. 
    
    Finally, we note that a class of 2D van der Waals antiferromagnets with chemical formula $M$P$X_3$ ($M$: transition metal, $X$: chalcogen) displaying zigzag antiferromagnetic order has been recently found to support an emergent 3-state Potts-nematic order parameter \cite{Wu2023,Hwangbo2024,Sun2024,Tan2024}. Future theoretical extensions of the present model to this case could unveil new physics, given the richer nematic domain landscape in the Potts-nematic phase compared to the Ising-nematic phase \cite{Chakraborty2023}, while the experimental techniques discussed above could be adapted to explore potential differences in dynamic scaling exponents and domain pattern formation between these distinct nematic systems.     

\begin{acknowledgments}
    YY and GWC were partially supported by the US Department of Energy Basic Energy Sciences under Contract No. DE-SC0020330. YHL is grateful for the support of Taiwan Ministry of Science and Technology Grant MOST: 110-2917-I-007-007. YY would like to thank W. J. Meese for intriguing discussion on related topics. The authors acknowledge Research Computing at The University of Virginia for providing computational resources and technical support that have contributed to the results reported within this publication. 
\end{acknowledgments}

\appendix
\setcounter{figure}{0}
\setcounter{equation}{0}
\renewcommand{\theequation}{A\arabic{equation}}
\renewcommand{\thefigure}{A\arabic{figure}}

\section{Correlation of domain at various $J_2/J_1$}\label{Sec:A1}

\begin{figure*}[h]
	   \includegraphics[width=\linewidth]{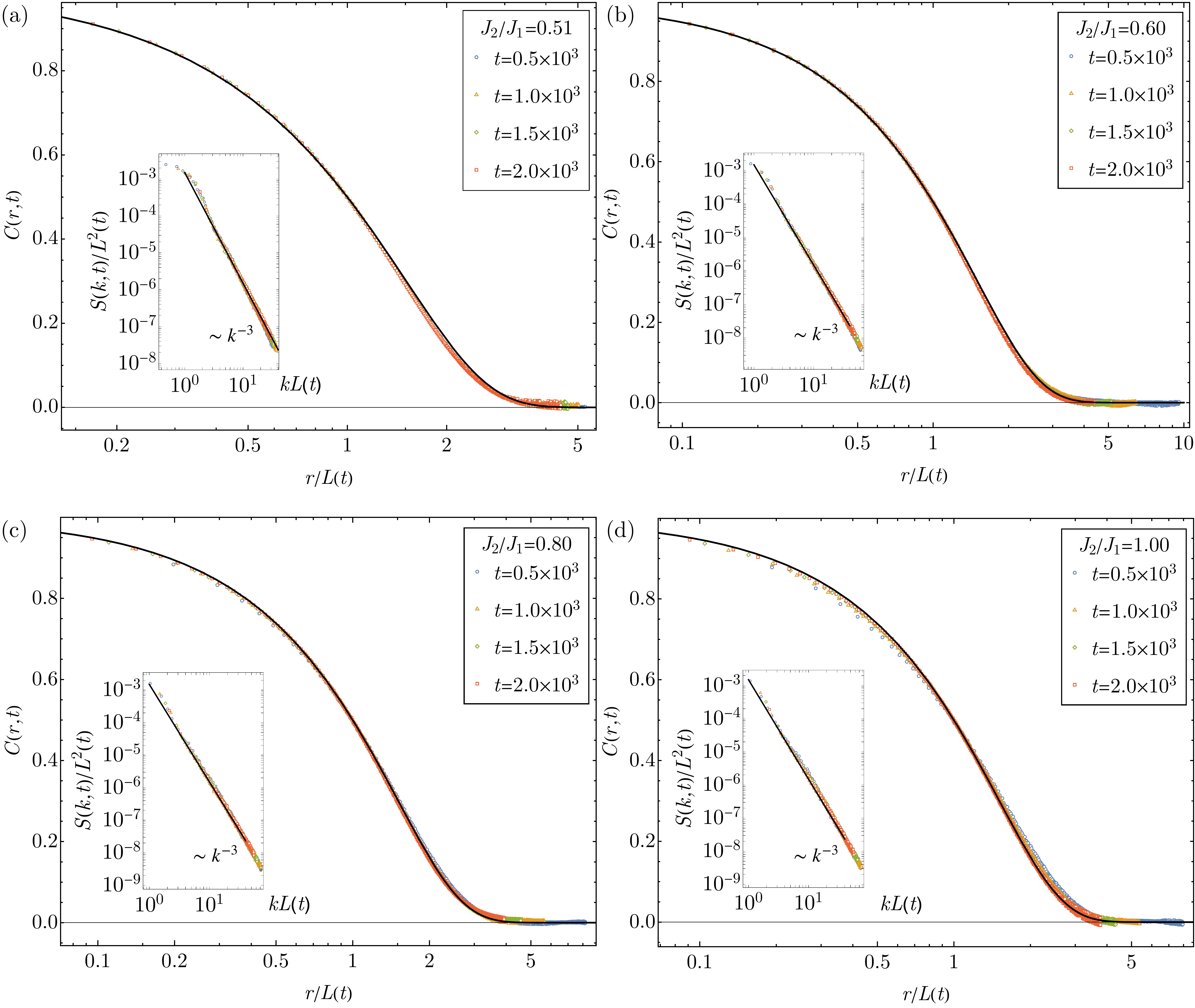}
	   \caption{Dynamic scaling for the equal-time correlation function $C(r,t)$ for (a) $J_2/J_1=0.51$, (b)  $J_2/J_1=0.60$,(c) $J_2/J_1=0.80$, and (d)  $J_2/J_1=1.00$ at $T/J_1=0.01$ in the linear--log scale for various times. The collapsed $C(r,t)$ is fitted with the OJK form of the correlation function $f_{\mathrm{OJK}}(\xi)$ (black solid line). The insets show that the structure factor tail follows the Porod's law $S(k)\sim k^{-3}$ in the log--log scale.}
	   \label{fig:A1} 
\end{figure*}

Here we present additional dynamic scaling results for the equal-time correlation function $C(r,t)$ across various $J_2/J_1$. We set the temperature sufficiently low ($T/J_1=0.01$) to provide us with a wide range choice of $J_2/J_1$, spanning from near the critical point ($J_2/J_1=0.51$) to deep within the nematic phase ($J_2/J_1=1.0$). The results are presented in Fig.~\ref{fig:A1}. The equal-time correlation function $C(r,t)$ all collapses into single curve after the rescaling, which can be described by the OJK form of the correlation function $f_{\mathrm{OJK}}(\zeta)$ given in Eq.~(\ref{eqn:OJK}). The Fourier transform of $C(r,t)$ displayed in the insets confirms the presence of a $k^{-3}$ power-law tail, consistent with Porod's law.

\section{Correlation of domain under weak disorder $W=0.03$ and $W=0.08$}\label{Sec:A2}

\begin{figure*}[h]
	   \includegraphics[width=\linewidth]{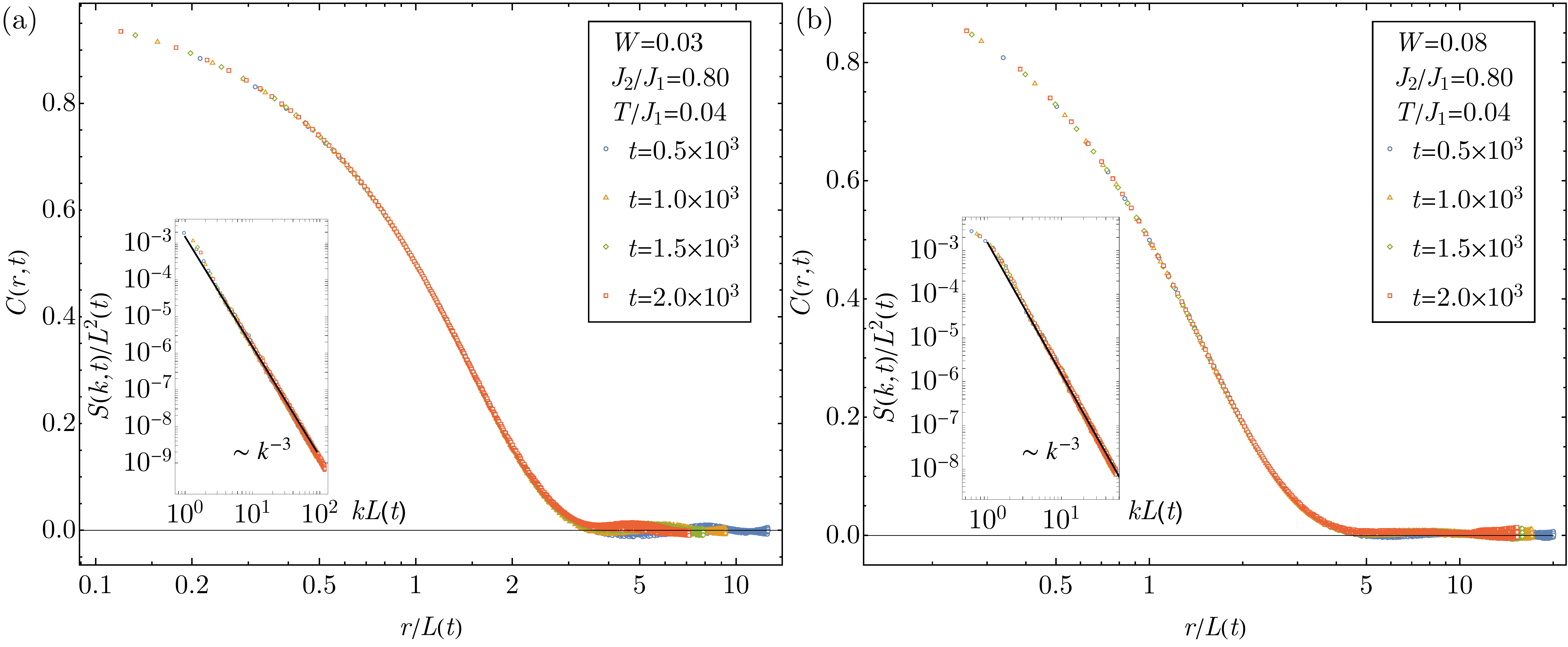}
	   \caption{Dynamic scaling for the equal-time correlation function $C(r,t)$ for $J_2/J_1=0.80$ and $T/J_1=0.04$ under the weak disorder strength (a) $W=0.03$ and (b) $W=0.08$ in the linear--log scale.The insets show that the structure factor tail follows the Porod's law $S(k)\sim k^{-3}$ in the log--log scale.}
	   \label{fig:A2} 
\end{figure*}

Here we present dynamic scaling results for the equal-time correlation function $C(r,t)$ under two weak disorder strength $W=0.03$ and $W=0.08$. The simulation parameters are identical to those used in Fig.~\ref{fig:9}(b). The results are presented in Fig.~\ref{fig:A2}. Correlation functions at both disorder strengths collapse onto a single curve, demonstrating dynamic scaling invariance is not affected by different weak disorder values. The insets of both plots show $k^{-3}$ power-law tail, in agreement with disorder-free scenarios.

\section{Saturation of local nematic order parameter}\label{Sec:A3}

\begin{figure*}[t]
	   \includegraphics[width=0.99\linewidth]{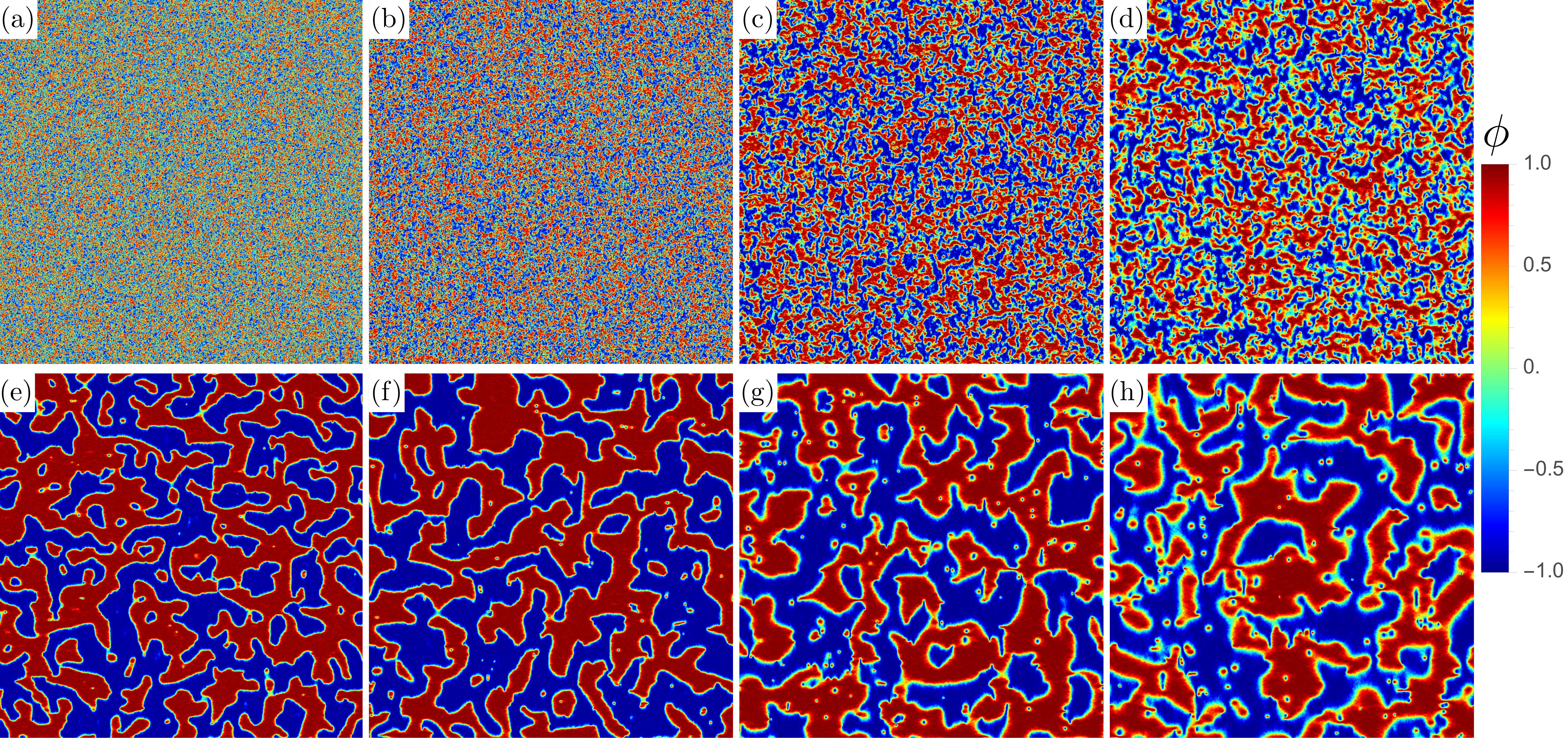}
	   \caption{The configuration of scalar nematic order parameter before reaching local equilibrium at $t=25$ for (a) $J_2/J_1=0.51$, (b) $J_2/J_1=0.60$, (c) $J_2/J_1=0.80$, (d) $J_2/J_1=1.00$, and after reaching local equilibrium at $t=200$ for (e) $J_2/J_1=0.51$, (f) $J_2/J_1=0.60$, (g) $J_2/J_1=0.80$, (h) $J_2/J_1=1.00$.}
	   \label{fig:A3} 
\end{figure*}

To verify the nematic order parameter fully saturates within the domain in the late-stage dynamics, we plot real space configuration in terms of the scalar local nematic order parameter defined in Eq.~(\ref{eqn:order_parameter_scalar}). Figure~\ref{fig:A3} compares configurations at $t=25$ (before $t_0$) and $t=200$ (after $t_0$) for various $J_2/J_1$ ratios. The top row reveals that before $t_0$, significant regions exhibit non-saturated order parameter values, particularly at smaller $J_2/J_1$. This aligns with our finding in Fig.~\ref{fig:6} that smaller $J_2/J_1$ values give larger $t_0$, indicating longer local equilibration times. The bottom row shows that for $t>t_0$, the scalar nematic order parameter saturates fully within each domain, with non-saturated values confined to narrow domain wall regions, confirming the transition to curvature-driven domain wall dynamics.


\nocite{*}
\bibliography{ref.bib}% Produces the bibliography via BibTeX.

\end{document}